\documentclass{ws-procs9x6}
\newcommand{\be}{\begin{equation}}
\newcommand{\ee}{\end{equation}}
\newcommand{\bea}{\begin{eqnarray}}
\newcommand{\eea}{\end{eqnarray}}
\newcommand{\fbBh}{\hat{f}_{b/B}}

\newcommand{\Mmpmphat}{\hat{M}_{-+;-+}^{0}}
\newcommand{\Mpppphat}{\hat{M}_{++;++}^{0}}

\newcommand{\Mmppmhat}{\hat{M}_{-+;+-}^{0}}

\newcommand{\phia}{\phi_{a}}

\newcommand{\dpcpphat}{\hat{D}^{\pi}_{++}}
\newcommand{\dpcpmhat}{\hat{D}^{\pi}_{+-}}

\newcommand{\phicc}{\phi^{\prime\prime}_c}
\newcommand{\xia}{\xi_{a}}
\newcommand{\xib}{\xi_{b}}
\newcommand{\xic}{\xi_{c}}

\newcommand{\xiat}{\tilde{\xi}_{a}}
\newcommand{\xibt}{\tilde{\xi}_{b}}
\newcommand{\xict}{\tilde{\xi}_{c}}
\newcommand{\xidt}{\tilde{\xi}_{d}}
\newcommand{\phihpi}{\phi_\pi^H}
\newcommand{\cal}{\mathcal}
\newcommand{\ud}{\mathrm{d}}
\newcommand{\Imm}{\mathrm{Im}}
\newcommand{\bkp}{\boldsymbol{k}_\perp}
\newcommand{\bkpa}{\boldsymbol{k}_{\perp a}}

\newcommand{\bfk}{\mbox{\boldmath $k$}}
\newcommand{\NP}[1]{{\it Nucl.\ Phys.}\ {\bf #1}}

\newcommand{\PL}[1]{{\it Phys.\ Lett.}\ {\bf #1}}
\newcommand{\PR}[1]{{\it Phys.\ Rev.}\ {\bf #1}}

\newcommand{\lax}{\lambda_{X}}

\newcommand{\laa}{\lambda_{a}}
\newcommand{\lab}{\lambda_{b}}
\newcommand{\lac}{\lambda_{c}}
\newcommand{\lad}{\lambda_{d}}
\newcommand{\la}{\lambda}

\def\lsim{\mathrel{\rlap{\lower4pt\hbox{\hskip1pt$\sim$}}\raise1pt\hbox{$<$}}}
\def\gsim{\mathrel{\rlap{\lower4pt\hbox{\hskip1pt$\sim$}}\raise1pt\hbox{$>$}}}
\def\nostrocostruttino#1\over#2{\mathrel{\mathop{\kern 0pt \rlap
{\hbox{$#1$}}} \hbox{\kern-.135em $#2$}}}
\def\sumint{\nostrocostruttino \sum \over {\displaystyle\int}}
\begin{document}

\title{A general formalism for single and double spin asymmetries
in inclusive hadron production
\footnote{\uppercase{T}alk delivered by \uppercase{S}.~\uppercase{M}elis
at the ``\uppercase{X} \uppercase{C}onvegno su 
\uppercase{P}roblemi di \uppercase{F}isica \uppercase{N}ucleare 
\uppercase{T}eo\-ri\-ca'', \uppercase{O}ctober 6-9, 2004, 
\uppercase{C}ortona, \uppercase{I}taly.}
}

\author{U. D'ALESIO\footnote{\uppercase{E}-mail: umberto.dalesio@ca.infn.it},
S. MELIS\footnote{\uppercase{E}-mail: stefano.melis@ca.infn.it},
F. MURGIA\footnote{\uppercase{E}-mail: francesco.murgia@ca.infn.it}}

\address{Istituto Nazionale di Fisica Nucleare, Sezione di Cagliari \\
and Dipartimento di Fisica, Universit\`a di Cagliari\\
C.P. 170,I-09042 Monserrato (CA),Italy\\}

\maketitle
 
\abstracts{We present a calculation of single and double spin asymmetries
for inclusive hadron production in hadronic collisions.
Our approach is based on Leading Order (LO) perturbative QCD and
generalized factorization theorems, with full account of intrinsic parton
momentum, $\bkp$, effects. This leads to a new class of spin and
$\bkp$-dependent distribution and fragmentation functions. Limiting
ourselves to consider leading twist functions, we show how they could 
play a relevant role in producing non-vanishing spin asymmetries.
}

\section{Introduction}
It has been experimentally known since a long time that transverse single
spin asymmetries (SSA) in hadronic collisions can be, in particular
kinematical regions, very large. Two relevant examples are: the
transverse $\Lambda$ polarization, $P_T^{\Lambda}$, measured in
unpolarized hadron collisions; the SSA, $A_{N}$, observed
in $p^{\uparrow}p\to\pi+X$. $P_T^{\Lambda}$ and $A_{N}$ are defined as:
\begin{equation} 
P_T^{\Lambda}=\frac{\ud\sigma^{AB\to \Lambda^{\uparrow} X}-
\ud\sigma^{AB\to \Lambda^{\downarrow} X}}
{\ud\sigma^{AB\to \Lambda^{\uparrow} X}+
\ud\sigma^{AB\to \Lambda^{\downarrow} X}}\quad
A_N=\frac{\ud\sigma^{A^{\uparrow}B\to C X}-
\ud\sigma^{A^{\downarrow}B\to C X}}
{\ud\sigma^{A^{\uparrow}B\to C X}+
\ud\sigma^{A^{\downarrow}B\to C X}}
\label{ptan}
\end{equation}
where $\ud\sigma$ stands for the corresponding invariant differential
cross section and $\uparrow, \downarrow$ denote transverse polarization
with respect to the hadron production plane.
Both these observables can reach in size values up to 30\%-40\%.
These results are at first puzzling in the context of perturbative QCD
(pQCD) if, as usual, one assumes a collinear partonic kinematics.
In fact, it is easy to see that in this case pQCD predicts almost
vanishing SSA at large energy scales.
Writing a transverse spin state as a combination of helicity states,
$|\uparrow/\downarrow\rangle=\frac{1}{\sqrt{2}}[|+\rangle\pm i|-\rangle]$,
we have schematically:
\begin{equation}
A_{N}\sim\frac{\langle\uparrow|\uparrow\rangle-\langle\downarrow|
\downarrow\rangle}{\langle\uparrow|\uparrow\rangle+
\langle\downarrow|\downarrow\rangle}
\sim\frac{2\Imm\langle+|-\rangle}{\langle+|+\rangle+\langle-|-\rangle}\,,
\label{anim} 
\end{equation}
where $\langle\uparrow|\uparrow\rangle$ stands for the partonic cross
section d$\hat\sigma^\uparrow$. 
The presence of a term like $\Imm \langle+|-\rangle$ in the numerator of
Eq.~(\ref{anim}) requires both helicity flip at the partonic level and a
relative phase between 
helicity amplitudes. Because massless QCD conserves helicity and Born level
amplitudes are real, $A_N$ should be proportional to the quark mass times 
a factor $\alpha_s$ coming from the required higher order contributions.
This implies 
$A_N\propto \alpha_s m_{q}/E_{q}$, which would be negligible at large energies.
A possible way out is to take into account partonic transverse momenta, $\bkp$,
 in parton distribution and fragmentation functions and in the
 elementary scattering process. 
This in turn leads to the introduction of a new class of spin 
and $\bkp$ dependent 
partonic distribution (PDF) and fragmentation (FF) functions that,
at least in principle, are able to generate spin asymmetries.
\section{General formalism}\label{sec:colkperp}
In the usual collinear pQCD approach at leading order (LO) and leading twist,
the unpolarized cross section for the process $AB \to CX$ reads:
\begin{eqnarray}\label{eq:collfac}
\ud\sigma^{AB \to CX} & = & 
\sum_{a,b,c,d} f_{a/A}(x_{a}, Q^{2})\otimes f_{b/B}(x_b, Q^2)\nonumber\\
& &\otimes\;\ud\hat\sigma^{ab \to cd}(\hat s, \hat t, \hat u, x_{a}, x_{b})
\otimes D_{C/c}(z, Q^{2})\,,
\end{eqnarray}
where $f_{a/A}(x_{a}, Q^{2})$ denotes the PDF for parton $a$
inside hadron $A$ carrying a fraction $x_{a}$ of the parent hadron 
light-cone momentum; 
$D_{C/c}(z, Q^{2})$ is the FF for parton $c$ fragmenting into hadron 
$C$ with a fraction $z$ of the parton light-cone 
momentum; $\ud\hat\sigma^{ab \to cd}$ 
denotes the partonic cross section for the elementary scattering $ab\to cd$;
$\hat s$, $\hat t$, $\hat u$ are the Mandelstam variables for the
partonic process.

In order to include $\bkp$ effects we have to 
generalize standard collinear PDF and FF to $\bkp$-dependent functions; 
for instance $f_{a/A}(x_{a})$ is generalized to $\hat{f}_{a/A}(x_{a},\bkpa)$,
where $\bkpa$ is the parton momentum perpendicular to the hadron
momentum, with the condition:
\begin{equation}
f_{a/A}(x_{a})=\int \ud^{2}\boldsymbol{k}_{\perp a} 
\hat{f}_{a/A}(x_{a},\boldsymbol{k}_{\perp a})\,.
\end{equation}
Analogously, the fragmentation function $D_{C/c}(z)$ is generalized to
$\hat D_{C/c}(z,\bfk_{\perp C})$, where $\bfk_{\perp C}$ is the transverse
momentum of the observed hadron $C$ with respect to the fragmenting parton
$c$ momentum.

The next step is to extend the pQCD expression for the cross section,
Eq.~(\ref{eq:collfac}), including $\bkp$ effects:
\begin{eqnarray}\label{eq:kpfac1}
\ud\sigma^{AB \to CX} &=& \sum_{a,b,c,d}\hat f_{a/A}(x_a,\bfk_{\perp
  a}; Q^2) \otimes \hat f_{b/B}(x_b, \bfk_{\perp b}; Q^2)\nonumber\\
&& \otimes\,\ud\hat\sigma^{ab \to cd}(\hat s, \hat t, \hat u, x_a, x_b)
\otimes \hat D_{C/c}(z, \bfk_{\perp C}; Q^2)\,,
\end{eqnarray}
where now $\hat s$, $\hat t$ and $\hat u$ depend on the full $\bkp$
kinematics.\cite{dm}  

A factorization theorem with the inclusion of transverse momenta has
not been formally proved in general,\cite{col} but only
for the Drell-Yan process, for two-particle inclusive production in
$e^+e^-$ annihilation\cite{css} and, recently, for SIDIS processes in
particular kinematical regions.\cite{ji}

In order to study spin asymmetries we have to include 
explicit spin dependences  into Eq.~(\ref{eq:kpfac1}). 
This can be done by introducing in the factorization scheme
the parton helicity density matrices
describing the parton spin states.
In this way, starting from Eq.~(\ref{eq:kpfac1})
we can write the polarized cross section as
\begin{eqnarray}\label{eq:polxsec1}
\ud\sigma^{(A,S_A) + (B,S_B) \to C + X}= \sum_{a,b,c,d, \{\la\}} 
\rho_{\la^{\,}_a,\la^{\prime}_a}^{a/A,S_A} \, 
\hat f_{a/A,S_A}(x_a,\bfk_{\perp a})\otimes&&\nonumber\\
\rho_{\la^{\,}_b, \la^{\prime}_b}^{b/B,S_B} \,
\hat f_{b/B,S_B}(x_b,\bfk_{\perp b})\otimes 
\hat M_{\la^{\,}_c, \la^{\,}_d; \la^{\,}_a, \la^{\,}_b} \,
\hat M^*_{\la^{\prime}_c, \la^{\,}_d; \la^{\prime}_a,\la^{\prime}_b}
\otimes 
\hat{D}^{\la^{\,}_C,\la^{\,}_C}_{\la^{\,}_c,\la^{\prime}_c}(z,\bfk_{\perp
  C}) &&
\end{eqnarray}
where $\{\la\}$ is a shorthand for the sum over all helicity indices involved.
In Eq.~(\ref{eq:polxsec1}),
$\rho_{\la^{\,}_a,\la^{\prime}_a}^{a/A,S_A}$ 
is the helicity density 
matrix of parton $a$ inside hadron $A$ with generic polarization $S_{A}$. 
Similarly for parton $b$ inside hadron $B$. 
The $\hat M_{\la^{\,}_c, \la^{\,}_d; \la^{\,}_a, \la^{\,}_b}$'s 
are the helicity amplitudes for the elementary process $ab\to cd$, 
normalized so that the unpolarized cross section, for a collinear
collision, is given by
\begin{equation}
\frac{d\hat\sigma^{ab \to cd}}{d\hat t} = \frac{1}{16\pi\hat s^2}\frac{1}{4}
\sum_{\la^{\,}_a, \la^{\,}_b, \la^{\,}_c, \la^{\,}_d}
|\hat M_{\la^{\,}_c, \la^{\,}_d; \la^{\,}_a, \la^{\,}_b}|^2\,.
\label{norm}
\end{equation}
The $\hat M_{\la^{\,}_c, \la^{\,}_d; \la^{\,}_a, \la^{\,}_b}$'s
in Eq. (\ref{eq:polxsec1})
are defined in the hadronic c.m. frame;
they are related to the usual helicity amplitudes, defined 
in the  ``canonical' partonic c.m. frame,
$\hat M_{\la^{\,}_c, \la^{\,}_d; \la^{\,}_a, \la^{\,}_b}^{0}$,
in a non trivial way by proper phases 
coming from the rotations and the boost connecting the two reference frames. 
These azimuthal phases are crucial in determining, when integrating
over partonic phase space,
the size of each allowed contribution to
Eq. (\ref{eq:polxsec1}).\cite{N1}${}^{,}$\cite{N2}${}^{,}$\cite{DMD} 
Finally,
$\hat D^{\la^{\,}_C,\la^{\prime}_C}_{\la^{\,}_c,\la^{\prime}_c}(z,
\bfk_{\perp C})$ is the product of {\it fragmentation amplitudes} for the
$c \to C + X$ process
\begin{equation}
\hat D^{\la^{\,}_C,\la^{\prime}_C}_{\la^{\,}_c,\la^{\prime}_c}
= \> \sumint_{X, \la_{X}} {\hat{\cal D}}_{\la^{\,}_{X},\la^{\,}_C;
\la^{\,}_c} \, {\hat{\cal D}}^*_{\la^{\,}_{X},\la^{\prime}_C;
\la^{\prime}_c}\,,
\label{framp}
\end{equation}
where $\sumint_{X, \la_{X}}$ stands for a spin sum and phase
space integration over all undetected particles, considered as a
system $X$. The usual unpolarized fragmentation function
$D_{C/c}(z)$, {\it i.e.} the number density
 of hadrons $C$ resulting from the fragmentation of an unpolarized
parton $c$ and carrying a light-cone momentum fraction $z$, is given by
\begin{equation}
D_{C/c}(z) = \frac{1}{2} \sum_{\la^{\,}_c,\la^{\,}_C} \int d^2\bfk_{\perp C}
\, \hat D^{\la^{\,}_C,\la^{\,}_C}_{\la^{\,}_c,\la^{\,}_c}(z, \bfk_{\perp C})
\,. \label{fr}
\end{equation}

One can also give a more explicit connection between 
hadron and parton polarizations. Interpreting the partonic distribution,
at LO, as the inclusive cross section for the process $A\to a+X$ we have:
\begin{eqnarray}
\rho_{\la^{\,}_a, \la^{\prime}_a}^{a/A,S_A} \>
\hat f_{a/A,S_A}(x_a,\bfk_{\perp a})
&=& \sum_{\la^{\,}_A, \la^{\prime}_A}
\rho_{\la^{\,}_A, \la^{\prime}_A}^{A,S_A}
\sumint_{X_A, \la_{X_A}} \!\!\!\!\!\!
{\hat{\cal F}}_{\la^{\,}_a, \la^{\,}_{X_A};
\la^{\,}_A} \, {\hat{\cal F}}^*_{\la^{\prime}_a,\la^{\,}_{X_A}; \la^{\prime}_A}
\label{distramp}\nonumber \\
&=&   \sum_{\la^{\,}_A, \la^{\prime}_A}
\rho_{\la^{\,}_A, \la^{\prime}_A}^{A,S_A} \>
\hat F_{\la^{\,}_A, \la^{\prime}_A}^{\la^{\,}_a,\la^{\prime}_a} \>,\label{defF}
\end{eqnarray}
where $\rho_{\la^{\,}_A, \la^{\prime}_A}^{A,S_A}$is the hadron 
helicity density matrix,
\begin{equation}
\hat{F}_{\la^{\,}_A, \la^{\prime}_A}^{\la^{\,}_a,\la^{\prime}_a} \equiv \>
\sumint_{X_A, \la_{X_A}} \!\!\!\!\!\!
{\hat{\cal F}}_{\la^{\,}_a,\la^{\,}_{X_A};\la^{\,}_A} \,
{\hat{\cal F}}^*_{\la^{\prime}_a,\la^{\,}_{X_A}; \la^{\prime}_A} \>,
\label{defFF}
\end{equation}
and the $\hat{\cal F}$'s are the {\it helicity distribution  
amplitudes} for the $A \to a + X$ process.
Inserting Eq.~(\ref{distramp}) into Eq.~(\ref{eq:polxsec1}) 
we obtain our master formula for a collision between polarized hadrons:
\begin{eqnarray}
\ud\sigma^{(A,S_A) + (B,S_B) \to C + X} &=&\!\!\!
\sum_{a,b,c,d, \{\la\}}\rho_{\la^{\,}_A, \la^{\prime}_A}^{A,S_A}
\>\hat F_{\la^{\,}_A, \la^{\prime}_A}^{\la^{\,}_a,\la^{\prime}_a}
\otimes\, \rho_{\la^{\,}_B, \la^{\prime}_B}^{B,S_B}
 \> \hat{F}_{\la^{\,}_B, \la^{\prime}_B}^{\la^{\,}_b,\la^{\prime}_b}
\nonumber \\
&& 
\otimes\,\hat M_{\la^{\,}_c, \la^{\,}_d; \la^{\,}_a, \la^{\,}_b}
\,\hat M^*_{\la^{\prime}_c, \la^{\,}_d;
\la^{\prime}_a,\la^{\prime}_b}\otimes
\hat{D}^{\la^{\,}_C,\la^{\,}_C}_{\la^{\,}_c,\la^{\prime}_c}\,.
\label{eq:master}
\end{eqnarray}
This expression contains all allowed combinations of spin
and $\bkp$-dependent distribution and fragmentation functions:
at LO and leading twist these functions have 
a simple partonic interpretation and are related 
to the spin and $\bkp$-dependent functions discussed
in other papers.\cite{bm}${}^{,}$\cite{bdr} 
As an example, the relation between the definitions of
the Sivers, the Boer-Mulders, and the Collins functions
in two widely adopted notations  
is the following:\cite{trento} 
\bea
\label{eq:Si}
f_{q/h^{\uparrow}}-f_{q/h^{\downarrow}} & \equiv &\Delta^{N}f_{q/h^{\uparrow}}=
4\Imm(\hat F_{+,-}^{+,+})\propto f_{1T}^{\perp} \\
\label{eq:BM}
f_{q^{\uparrow}/h}-f_{q^{\downarrow}/h} & \equiv &\Delta^{N}f_{q^{\uparrow}/h}=
2\Imm(\hat F_{+,+}^{+,-})\propto h_{1}^{\perp} \\
\label{eq:Co}
D_{h/q^{\uparrow}}-D_{h/q^{\downarrow}} &\equiv &\Delta^{N}D_{h/q^{\uparrow}}=
2\Imm(D^{h}_{+-})\propto H_{1}^{\perp}\,.
\eea
In Eq.~(\ref{eq:Si}) $\Delta^{N}f_{q/h^{\uparrow}}$ 
gives the probability to find an unpolarized 
parton inside a transversely polarized hadron.
$ \Delta^{N}f_{q^{\uparrow}/h}$ in Eq.~(\ref{eq:BM}),
is the probability to find 
a transversely polarized quark inside an unpolarized hadron. 
Finally, $\Delta^{N}D_{h/q^{\uparrow}}$ is the probability for 
a transversely polarized quark to fragment into an unpolarized hadron.

\subsection{$A_N(A^\uparrow\!\!B\to \pi+X$)}
As an application of this formalism let us consider the process 
$A^{\uparrow}B\to \pi+X$,
where $A$ and $B$ are spin one-half hadrons. 
In this case the helicity density matrices for $A$, $B$, take the form:
\begin{equation}
\rho_{\la^{\,}_A, \la^{\prime}_A}^{A,\uparrow/\downarrow}=
\frac{1}{2}\left(
\begin{array}{cc}
 1 & \mp i \\
 \pm i &  1\\
\end{array}
\right) \quad\quad\quad
\rho_{\la^{\,}_B,\la^{\prime}_B}^{B,0} = \frac{1}{2}
\left(
\begin{array}{cc}
 1 & 0 \\
 0 & 1 \\
\end{array}
\right)\,.\label{eq:rho}
\end{equation}
By performing explicitly the sum over  hadron helicity 
indices in Eq.~(\ref{eq:master}) 
and using Eq.~(\ref{eq:rho})  we obtain:
\begin{eqnarray}
&&\ud\sigma^{A^{\uparrow/\downarrow} + B \to \pi + X} =\!\!\!
\sum_{a,b,c,d,\{\la\}}\!\!\![\hat F_{+,+}^{\laa,\laa'} + 
\hat F_{-,-}^{\laa,\laa'}\mp i(\hat F_{+,-}^{\laa,\laa'}-
\hat F_{-,+}^{\laa,\laa'})]\nonumber\\
&&\qquad\qquad\otimes\,
[\hat F_{+,+}^{\lab,\lab'}+\hat F_{-,-}^{\lab,\lab'}]\otimes
\hat{M}_{\lac\lad;\laa\lab}
\hat{M}^{*}_{\lac^{'}\lad;\laa^{'}\lab^{'}} 
\otimes \hat{D}^{\pi}_{\lac,\lac'}\,,
\label{mastertra}\end{eqnarray}
where $\{\la\}$ stands for a sum over partonic helicity indices.

In  Eq.~(\ref{mastertra}) there are terms 
that change their sign when changing the sign of the
corresponding hadron polarization. 
These terms survive in the numerator of the asymmetry and 
involve, depending also on the partonic subprocess considered, different
combinations of the Sivers, Boer-Mulders and Collins mechanisms. 

As an explicit example let us now consider a particular
partonic subprocess: $qq\to qq$. 
Then, using known symmetry properties of the helicity distribution
functions\cite{N1} we can write:
\bea
\hat F_{\la^{\,}_A,\la^{\prime}_A}^{\la^{\,}_a,\la^{\prime}_a}
(x_a, \bfk_{\perp a})
& = & 
F_{\la^{\,}_A,\la^{\prime}_A}^{\la^{\,}_a,\la^{\prime}_a}(x_a, k_{\perp a})
\> {\rm exp}[i(\la^{\,}_A - \la^{\prime}_A)\phi_a] \>, \label{fft-ff} \\
{F}_{-\la^{\,}_{A},-\la^{\prime}_A}^{ -\la^{\,}_a,-\la^{\prime}_{a}}
& = & 
(-1)^{(\la^{\,}_A -\la^{\,}_a) + (\la^{\prime}_A -\la^{\prime}_a)} \>
{F}_{\la^{\,}_{A},\la^{\prime}_A}^{ \la^{\,}_a,\la^{\prime}_{a}}
\,,
\label{parFF}
\eea
where $\phi_{i}$ ($i=a,b,c,d$) are the azimuthal 
angles of parton $i$ three-momentum in the $AB$ c.m. frame. 
Analogously for fragmentation functions we have:
\be 
{\hat D}_{\la^{\,}_{c},\la^{\prime}_c}^{\pi}
={D}_{\la^{\,}_{c},\la^{\prime}_c}^{\pi} \> {\rm
exp}[i(\la^{\,}_{c} - \la^{\prime}_c)\phi_\pi^H] \>,
\label{ddt-dd} 
\ee
where $\phi_\pi^H$ is the azimuthal angle of the pion three-momentum
as seen in the parton $c$ helicity frame.\cite{N1} 

As mentioned above,
the helicity amplitudes $\hat{M}_{\lac\lad;\laa\lab}$
in Eq. (\ref{mastertra}) are defined in the hadronic c.m. frame;
we can relate these amplitudes to those given in the canonical
partonic c.m. frame,  
$\hat{M}^{0}_{\lac\lad;\laa\lab}$, where $Z$ is the direction of the colliding 
partons and the $XZ$-plane coincides with the scattering plane. 
In this frame the $\hat{M}^{0}_{\lac\lad;\laa\lab}$'s exhibit explicit 
symmetry properties. \\
To reach the canonical partonic c.m. frame, from 
the hadronic c.m. frame,
we have first to perform a boost along the
$\boldsymbol{p}_{a}+\boldsymbol{p}_{b}$ direction,
so that the boosted three-vector  
$\boldsymbol{p}_{a}^{\prime}+\boldsymbol{p}_{b}^{\prime}$ is vanishing.
In this frame ($S'$) partons $a$ and $b$ are in the ``head-on'' configuration,
but not aligned along the $Z$-axis direction.
We then perform a rotation to align the colliding initial partons
with the $Z$-axis. 
We call this new frame $S^{\prime\prime}$. In this frame parton $c$ 
three-momentum does not lie in the $XZ$-plane 
but has a transverse component, with an azimuthal angle $\phicc$. 
A final rotation around $Z$ by $\phicc$ leads to the canonical configuration.
The relationship between 
$\hat{M}_{\lac\lad;\laa\lab}$ and $\hat{M}^{0}_{\lac\lad;\laa\lab}$
is then given by:\cite{N1}
\bea
&&\hat M_{\la^{\,}_c, \la^{\,}_d; \la^{\,}_a, \la^{\,}_b} =
 \hat M^0_{\la^{\,}_c, \la^{\,}_d; \la^{\,}_a, \la^{\,}_b} \nonumber\\
&&\qquad\qquad
\times e^{-i [\la^{\,}_a (\xi _a+\xiat) + \la^{\,}_b (\xi _b+\xibt) -
          \la^{\,}_c (\xi _c+\xict) - \la^{\,}_d (\xi _d+\xidt)]}
\,  e^{i(\la^{\,}_a - \la^{\,}_b)\phi^{\prime\prime}_c}\\
&&\qquad\qquad\quad\;\: 
= \hat M^0_{\la^{\,}_c, \la^{\,}_d; \la^{\,}_a, \la^{\,}_b} \nonumber\\
&&\qquad\qquad\times e^{-i (\la^{\,}_a \xi _a + \la^{\,}_b \xi _b -
          \la^{\,}_c \xi _c - \la^{\,}_d \xi _d)}
\,  e^{-i [(\la^{\,}_a - \la^{\,}_b) \tilde \xi _a -
         (\la^{\,}_c - \la^{\,}_d) \tilde \xi _c]}
\, e^{i(\la^{\,}_a - \la^{\,}_b)\phi^{\prime\prime}_c}\,,\nonumber
\label{M-M0}
\eea
where $\xi_i$, $\tilde{\xi}_{i}$ ($i=a,b,c,d$) are phases due to
the behaviour of helicity states 
under the Lorentz transformations connecting the hadronic 
and the canonical partonic c.m. 
frame.\cite{N1}${}^{,}$\cite{elliot}

Using Eqs.~(\ref{fft-ff}), (\ref{parFF}), (\ref{ddt-dd}), (\ref{M-M0})
into Eq.~(\ref{mastertra})
we obtain the contribution to the numerator of $A_{N}$
coming from the $qq\to qq$ partonic subprocess:
\bea
&& \ud \sigma^{A^{\uparrow}B\rightarrow\pi X}_{qq\to qq}-
\ud \sigma^{A^{\downarrow}B\rightarrow\pi X}_{qq\to qq}
\propto\nonumber\\
&&\;\;\;\;2\:\Imm[\hat{F}_{+-}^{++}(x_a,k_{\perp a})]\cos\phia
\otimes\fbBh(x_b,k_{\perp b})\nonumber\\
&&\qquad\qquad\otimes\left(|\Mpppphat|^{2}+
|\Mmpmphat|^{2}+|\Mmppmhat|^{2}\right)
\otimes\dpcpphat\nonumber\\
&& -\left\{\hat{F}_{+-}^{+-}(x_a,k_{\perp a})
\cos(2\phicc-\xia-\xiat +\xib+\xibt +\phia)\right\}\nonumber\\
&&\qquad\left.-\;\hat{F}_{+-}^{-+}(x_a,k_{\perp a})
\cos(2\phicc-\xia-\xiat +\xib+\xibt -\phia)\right\}\nonumber\\
&&\qquad\qquad\otimes\;4\:\Imm[\hat{F}_{++}^{+-}(x_b,k_{\perp b})]
\otimes \Mmppmhat\Mmpmphat \otimes \dpcpphat \\
&& + \left\{\hat{F}_{+-}^{+-}(x_a,k_{\perp a})
\cos(\phicc+\phihpi-\xia-\xiat+\xic+\xict+\phia)\right. \nonumber\\
&&\qquad-\left.\hat{F}_{+-}^{-+}(x_a,k_{\perp a})
\cos(\phicc+\phihpi-\xia-\xiat+\xic+\xict-\phia)\right\}\nonumber\\
&&\qquad\qquad\otimes\,\fbBh(x_b,k_{\perp b})\otimes
\Mpppphat\Mmpmphat\otimes 2\, \Imm(\dpcpmhat)\nonumber\\
&& - \;8\;\Imm[\hat{F}_{+-}^{++}(x_a,k_{\perp a})]
\cos\phia\otimes\Imm[\hat{F}_{++}^{+-}(x_b,k_{\perp b})]\nonumber\\
&&\qquad\times \cos(\phicc-\phihpi+\xib+\xibt-\xic-\xict) 
\otimes\Mpppphat\Mmppmhat\otimes\Imm(\dpcpmhat)\nonumber\,.
\eea
With the help of Eqs.~(\ref{eq:Si}), (\ref{eq:BM}) and (\ref{eq:Co}) one can
identify in this expression the Sivers, Boer-Mulders and Collins
mechanisms. 
The terms $\hat{F}_{+-}^{+-}$ and $\hat{F}_{+-}^{-+}$ are in turn
related to the distribution of transversely polarized quarks inside a 
transversely polarized hadron, the well-known transversity function.

\section{Final hadron polarization in hadronic collisions }
Within the same formalism we are able to calculate polarized cross sections 
for processes in which the final, spin one-half hadron (e.g. a
$\Lambda$ hyperon), is polarized.  
Our master formula, Eq.~(\ref{eq:master}), is modified 
by the introduction of the  helicity density matrix 
$\rho_{\lambda_{\Lambda},\lambda_{\Lambda}'}^{\Lambda}$ 
of the observed hadron, describing its polarization state. 
This way, Eq.~(\ref{eq:master}) becomes:
\begin{eqnarray}
\label{rogen}
&&
\rho_{\lambda_{\Lambda},\lambda_{\Lambda}'}^{\Lambda}
\ud \sigma^{(A,S_A) (B,S_B)\to \Lambda X} 
 \propto \,
\sumint_{\lax,X}\sum_{\lac,\lac'}
{\hat{\cal D}}_{\lambda_{\Lambda},\lax,\lac}
{\hat{\cal D}}^{*}_{\lambda_{\Lambda}',\lax,\lac'}
\rho_{\lac,\lac'}^{\Lambda/c}\nonumber\\
&&=\!\!\!\!\!\!\!\sum_{a,b,c,d, \{\la\}}\!\!\!\!\!
\rho_{\la^{\,}_A, \la^{\prime}_A}^{A,S_A} \,
\hat F_{\la^{\,}_A, \la^{\prime}_A}^{\la^{\,}_a,\la^{\prime}_a}
\otimes \rho_{\la^{\,}_B, \la^{\prime}_B}^{B,S_B} 
\, \hat{F}_{\la^{\,}_B, \la^{\prime}_B}^{\la^{\,}_b,\la^{\prime}_b} 
 \otimes\,\hat{M}_{\lac\lad;\laa\lab} 
 \hat{M}^{*}_{\lac^{'}\lad;\laa^{'}\lab^{'}} 
\otimes
\hat{D}_{\lac,\lac'}^{\lambda_{\Lambda},\lambda_{\Lambda}'}\,.\nonumber\\ 
&&
\end{eqnarray}
By choosing $S_A$ and $S_B$, and performing the sum over partonic
helicity indices (as discussed above),   
Eq.~(\ref{rogen}) allows one to compute all polarization states
(longitudinal and transverse) for 
a final spin-1/2 hadron produced in (un)polarized hadron-hadron collisions.
Notice also the appearance in the fragmentation sector of new terms with
respect to the pion case, depending on the final hadron helicities.

\section{Conclusions}
Spin effects in inclusive high-energy 
hadronic reactions play an important role in
our understanding of strong interactions.\\ 
We have presented a general approach to describe, within pQCD  
factorization schemes and using the helicity formalism,
polarized inclusive particle production in 
high-energy hadronic collisions.
By taking into account intrinsic motion 
of partons in the distribution and fragmentation functions and in the 
partonic process, this approach
allows one to give explicit expressions for
single and double spin asymmetries.
This requires the introduction of a new class of
spin and $\bkp$-dependent functions.
The combined study of single and double spin asymmetries for
different particles and in different kinematical situations
may help in gathering information on these basically unknown
functions.
As an example, we have briefly discussed two interesting
applications of this approach, namely $A_N(A^\uparrow B\to \pi+X)$
and final hadron polarization in (un)polarized hadronic collisions. 



\begin{thebibliography}{0}
\bibitem{dm}
  U. D'Alesio and F. Murgia, \PR{D70} (2004) 074009
\bibitem{col}
  J.C. Collins, \NP {B396} (1993) 161
\bibitem{css}
  J.C. Collins, D.E. Soper and G. Sterman, \NP {B250} (1985) 199;
  J.C. Collins and D.E. Soper, \NP {B193} (1981) 381
\bibitem{ji}
  X. Ji, J.-P. Ma and F. Yuan, e-Print Archive: hep-ph/0404183;
  \PL{B597} (2004) 299
\bibitem{N1}
  M. Anselmino, M. Boglione, U. D'Alesio, E. Leader and F. Murgia 
  e-Print Archive: hep-ph/0408356 ({\it Phys. Rev. D}, in press)
\bibitem{N2}
  M. Anselmino, M. Boglione, U. D'Alesio, E. Leader, S. Melis and 
  F. Murgia in preparation
\bibitem{DMD}
M. Anselmino, M. Boglione, U. D'Alesio, E. Leader and F. Murgia,
 \PR {D70} (2004) 074025
\bibitem{bm}
  D. Boer, P. Mulders and F. Pijlman, \NP{B667} (2003) 201
\bibitem{bdr}
  V. Barone, A. Drago and P. Ratcliffe, {\it Phys. Rep.} {\bf 359} (2002) 1
\bibitem{trento}
 A. Bacchetta, U. D'Alesio, M. Diehl, C. Andy Miller,
 e-Print Archive: hep-ph/0410050 ({\it Phys. Rev. D}, in press)
\bibitem{elliot}
  For a pedagogical introduction to all the basics of helicity formalism, see,
  {\it e.g.}, E. Leader, {\it Spin in Particle Physics}, Cambridge University
  Press, 2001


\end{thebibliography}
\end{document}